\documentclass[10pt,A4,conference]{IEEEtran}

\usepackage{geometry}
 \usepackage{authblk}           
\usepackage[T1]{fontenc}
\usepackage{cite}
\usepackage{balance}
\usepackage[export]{adjustbox}

\ifCLASSOPTIONcompsoc
 \usepackage[caption=false,font=normalsize,labelfont=sf,textfont=sf]{subfig}
\else
 \usepackage[caption=false,font=footnotesize]{subfig}
\fi

\usepackage{algorithm}
\usepackage[noend]{algpseudocode}                       
\usepackage{amssymb}
\usepackage{amsthm}
\usepackage{array}
\usepackage{graphicx}
\usepackage{lettrine}
\usepackage{epsf} 
\usepackage{psfrag}
\usepackage[usenames,dvipsnames]{pstricks}
\usepackage{pst-grad} 
\usepackage{pst-plot} 
\usepackage[space]{grffile} 
\usepackage{etoolbox} 
\usepackage{url} 

\usepackage{amsmath}

\begin{document}

\title{Impact of Reactive Jamming Attacks on LoRaWAN: a Theoretical and Experimental Study}

\author[]{Amavi Dossa$^*$}
\author[]{Andreas Burg$^\circ$}
\author[]{El Mehdi Amhoud$^*$}
\affil[]{$^*$College of Computing, Mohammed VI Polytechnic University (UM6P), Benguerir, Morocco, \authorcr
$^\circ$ Telecommunication Circuits Laboratory, Ecole polytechnique fédérale de Lausanne, Switzerland
}

\maketitle

\begin{abstract}
This paper investigates the impact of reactive jamming on LoRaWAN networks, focusing on showing that LoRaWAN communications can be effectively disrupted with minimal jammer exposure time. The susceptibility of LoRa to jamming is assessed through a theoretical study of how the frame success rate is impacted by only a few jamming symbols. Different jamming approaches are studied, among which repeated-symbol jamming appears to be the most disruptive, with sufficient jamming power. A key contribution of this work is the proposal of a software-defined radio (SDR)-based jamming approach implemented on GNU Radio that generates a controlled number of random symbols, independent of the standard LoRa frame structure. This approach enables precise control over jammer exposure time and provides flexibility in studying the effect of jamming symbols on network performance. The theoretical analysis is validated through experimental results, where the implemented jammer is used to assess the impact of jamming under various configurations. Our findings demonstrate that LoRa-based networks can be disrupted with a minimal number of symbols, emphasizing the need for future research on stealthy communication techniques to counter such jamming attacks.  
\end{abstract}

\IEEEpeerreviewmaketitle

\section{Introduction}

The internet of things (IoT) has transformed numerous sectors, including healthcare, agriculture, and smart cities, by enabling seamless communication and data sharing among interconnected devices. At the heart of IoT systems lies the communication protocol, which is essential for data transfer between devices. IoT networks consist of resource-constrained devices that must communicate over long distances. However, most existing communication technologies are either limited in range, such as Zigbee and Bluetooth, or energy-hungry, like 802.11x \cite{ayoub2018internet_lorawan_bluetooth_zigbee_wifi}. This led to the emergence of low power wide area network (LPWAN) technologies, including long range (LoRa) and its network protocol LoRaWAN (long range wide area network), which have gained popularity due to their long-range capability and cost-effectiveness.

LoRaWAN's increasing adoption has led to several challenges, particularly related to scalability and security \cite{jouhari_survey_scalable_lorawan}. The growing number of devices increases packet collisions and reduces the network capacity due to the uncoordinated Aloha multiple access scheme used by LoRaWAN. Various solutions have been proposed to enhance network capacity, such as the adoption of slotted Aloha \cite{dossa2024duty_cycle_efficient, polonelli2019slotted_aloha_on_lorawan}, support for carrier sensing \cite{gamage2023lmac_efficient_carrier_sense_lora}, and multi-user receivers with interference cancellation at the gateway \cite{9723151_tapparel_interference_cancellation_lora}. Regarding security, several attacks have been identified, including message replay targeting node activation, packet forging \cite{avoine2018rescuing_lorawan_1.0, Coman2019_security_issues_iot_lorawan_sigfox_nbiot}, battery exhaustion, and denial-of-service (DoS) attacks \cite{Mikhaylov2019_energy_attack}, along with corresponding countermeasures \cite{Prasad2022_analysis_prevention_dos_attacks_lorawan}.

Among these, DoS attacks from jamming are particularly problematic for the network. They are difficult to fully mitigate and easy to execute, especially since LoRa operates in the ISM (industrial, scientific, and medical) band. However, their impact remains underexplored in the literature. Previous studies on LoRa jamming typically used static setups, where the jammer was configured with a fixed frame length and power solely to demonstrate the effectiveness of jamming.

In this paper, we investigate the impact of jamming on LoRaWAN networks under various configurations. We demonstrate that LoRa-based networks can be jammed with a minimal number of symbols, effectively blocking their traffic, and identify the exact number of symbols required for this disruption. Given that jamming is facilitated by the easily detectable LoRa preamble, this study highlights the need for further research into covert communication techniques for LoRa, especially for critical nodes. Our contributions are as follows: 
\begin{itemize} 
    \item We analyze how LoRa modulation behaves under jamming and derive expressions to quantify the impact of jamming on frame success rates. Specifically, the derived formulas describe how the number of jamming symbols affects the frame success rate.
    \item We compared several jamming approaches, depending on the jamming signal length and power required to disrupt the network. 
    \item We propose a software defined radio (SDR)-based jamming approach that generates a controlled number of random symbols, independent of the standard LoRa frame structure, that enables precise control over jammer exposure time and facilitates experiments on the effects of jamming on the network.
\end{itemize}

The rest of the paper is organized as follows: Section II reviews related work on LoRa jamming. Section III provides an overview of the LoRa modulation. Section IV analyzes the LoRa modulation behavior under jamming and introduces our jamming framework. Section V presents the experimental validation of our analysis, and Section VI concludes the paper and suggests future work.



\section{Related Work}
The increasing adoption of LoRaWAN has triggered growing concerns about its security, prompting a series of assessments across the different layers of its communication stack. Among these, physical layer jamming attacks have received particular attention due to their ease of implementation and difficulty in mitigation. Such attacks pose a significant threat, potentially resulting in partial or complete denial of service (DoS). 

\cite{nafees2020impact_of_dos_attacks_on_energy_consumption} explores the impact of DoS attacks on the energy consumption of LoRaWAN end-devices. Jamming is also the basis of energy depletion attacks \cite{Mikhaylov2019_energy_attack}, where an attacker first jams the uplink transmission and then impersonates the gateway through a downlink transmission in RX2 to keep the device active. \cite{Pirayesh2021_jamming_anti-jamming_wireless_networks_survey} provides a detailed analysis of two jamming variants: reactive and selective jamming. In reactive jamming, the attacker listens to the channel and only jams when a frame is detected, whereas selective jamming targets specific devices, leaving others unaffected. \cite{martinez2019performance_lorawan_jamming} investigates the performance of LoRaWAN networks under jamming conditions, evaluating scenarios with and without node retransmission. The study used a network simulator, and the jammer in this case did not listen to the channel, instead randomly transmitting jamming frames.

Several methods have been proposed to detect and/or mitigate jamming attacks on LoRa/LoRaWAN, such as exploiting the differences between jamming signals and legitimate transmissions in the power domain to distinguish between the two \cite{Hou2022_jamming_lora_phy_and_countermeasure}.

Most of these studies follow a similar experimental setup: a jammer based on a conventional LoRa chip transmits a complete LoRa frame to interfere with ongoing transmissions, increasing the air-time of the jamming signal and the likelihood of detection. However, the effects of jamming frame length and its impact on the inherent LoRa protection mechanisms have not been thoroughly investigated in the literature. In this paper, we propose to reduce the number of jamming symbols as much as possible based on the transmission success rate in the presence of a given number of jamming symbols.

\section{Overview of LoRa Modulation}
LoRa is a proprietary modulation technique that employs chirp spread spectrum (CSS) to achieve long-range communication at low power and low data rate. It defines a spreading factor (SF) that determines symbol duration and bit width.

The discrete-time baseband expression of a LoRa-modulated chirp, sampled at the Nyquist frequency, is given by \cite{ghanaatian2019lora_baseband}.
\begin{equation}
    x_S[n] = e^{j2\pi \Big(\frac{n^2}{2N} + \big(\frac{S}{N} - \frac{1}{2} - u\big[n-n_f\big]\big)n \Big)} \mbox{,}
    \label{eq:lora_baseband_exp_compactd}
\end{equation}
with $S$ the modulated symbol, $N=2^{SF}$ the number of chips, $n_f = N - S$ the folding frequency in discrete time.

The signal received by a receiver can be expressed as
\begin{equation}
    y[n] = h x_S[n] + z[n] \mbox{,}
\end{equation}

where \( h \) is the channel gain and \( z[n] \) is the complex-valued white Gaussian noise with zero mean and variance \( \sigma^2 \). The demodulation process consists of three main steps. First, a dechirping operation is performed, where \( y[n] \) is multiplied by \( x_0^* \), a downchirp. This is followed by a discrete Fourier transform (DFT):
\begin{equation}
    Y[k] = \text{DFT}\left\{ y[n] x_0^*[n] \right\}
\end{equation}
Finally, the index of the bin with the highest energy is taken as the demodulated symbol:
\begin{equation}
    \begin{matrix}
        \hat S = \mbox{argmax}(|Y[k]|)  \\
        k
    \end{matrix}
\end{equation}
\begin{figure}[!t]
    \centering
    \includegraphics[width=\linewidth]{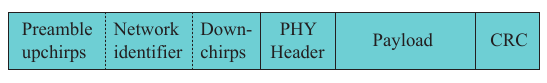}
    \caption{LoRa packet structure}
    \label{fig:lora-packet-structure}
    \vspace{-15pt}    
\end{figure}
LoRa defines a PHY frame structure, as shown in \mbox{Fig. \ref{fig:lora-packet-structure}}. The optional header includes the payload length and coding rate and is mandatory for LoRaWAN frames. The optional 16-bit CRC field is required for uplink LoRaWAN frames but is not used for downlink.

The preamble upchirps mark the beginning of a LoRa frame. The number of upchirps is configurable between 6 and 65,535 (8 for LoRaWAN), and the downchirps are 2.25 symbols long. The network identifier, which is 2 symbols long, distinguishes frames from different networks. Therefore, the minimum length of the LoRa preamble is 10.25 symbols. Consequently, a jammer based on a conventional LoRa chip produces a jamming signal that is more likely to expose the jammer. To resolve this issue, we propose to limit the jamming to only a few
LoRa symbols, just sufficient to disrupt the decoding of the original frame
with sufficiently high probability.

\section{Jamming Analysis Against LoRa protection Mechanisms}

\begin{figure}[!t]
    \centering
    \includegraphics[width=\linewidth]{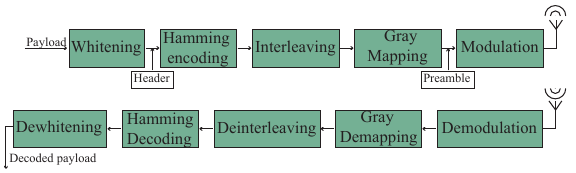}
    \caption{LoRa transceiver chain}
    \label{fig:lora-transceiver-chain}
    \vspace{-10pt}    
\end{figure}

\begin{figure}[!t]
    \centering
    \subfloat[]{\includegraphics[width=0.45\linewidth, valign=c]{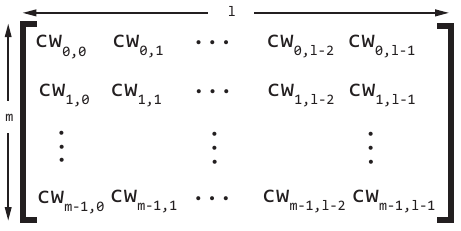}}
    \hfill
    \subfloat[]{\includegraphics[width=0.45\linewidth, valign=c]{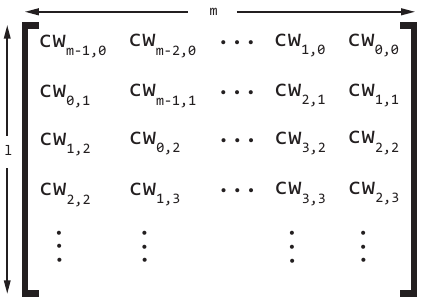}}
    \caption{LoRa diagonal interleaving: (a) before, (b) after; $m=$ SF, $l=4+$ CR}
    \label{fig:interleaving-general}
\end{figure}

    

\begin{figure}[!t]
    \centering
    \includegraphics[width=0.8\linewidth]{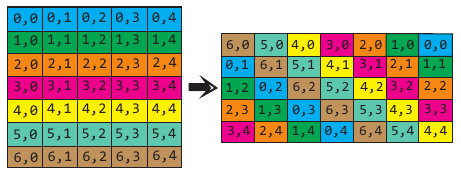}        
    \caption{LoRa interleaving example with SF 7 and CR 1}
    \label{fig:interleaving-sf7-r1}
    \vspace{-15pt}    
\end{figure}



In this section, we review the mechanisms implemented in LoRa against communication errors and analyze the impact of jamming on the decoding of received frames with respect to these mechanisms.

For a jamming attack to be effective, the jamming signal must be difficult for the LoRa receiver to mitigate. Several studies have examined the impact of various interference sources on LoRa \cite{courjault2020robust_lora_against_high_impulsive_noise, mikhaylov2017lorawan_scalability_susceptibility_internetwork_interference, haxhibeqiri2018sub_gigahertz_interference_harmful_for_lora}. These studies show that LoRa transmissions are most affected by interference from another LoRa transmission using the same SF on the same channel, referred to as intra-SF/same-SF interference.

The LoRa transceiver chain, as shown in Fig. \ref{fig:lora-transceiver-chain}, consists of several components that work together to detect and correct transmission errors. The Hamming encoding/decoding stage provides initial error protection through forward error correction (FEC) codes. These codes are associated with four coding rates (CR): 1, 2, 3, and 4, which correspond to Hamming codes (4,5), (4,6), (4,7), and (4,8), respectively. CR 1 and 2 can only detect errors, while CR 3 and 4 can correct a single-bit error per codeword. This protection is further enhanced by the interleaving/deinterleaving stage. The interleaving process spreads the bits of a codeword across multiple symbols, which, in combination with Hamming encoding, strengthens the frame's resilience to burst errors.

The interleaving stage operates as a block transformation. It takes $m$ SF codewords, each of length $l = 4 +$ CR bits, output by the Hamming encoding block, and produces $l$ symbols of $m$ bits. Reverse engineering has revealed that LoRa uses a diagonal interleaving scheme, shown in Fig. \ref{fig:interleaving-general}. The input to the interleaver is an $m \times l$ matrix, where each row contains the bits of a codeword. The output is an $l \times m$ matrix, where the $i$-th row corresponds to the $j$-th column of the input matrix, starting with the last bit and proceeding downwards, followed by an $i$-fold right rotation. The interleaved output symbols contain exactly one bit from each input symbol. An example of this diagonal interleaving for SF 7 and CR 1 is shown in Fig. \ref{fig:interleaving-sf7-r1}.
 
A reactive jamming scenario is considered, where the jamming signal is assumed to have sufficient power to disrupt the legitimate signal at the receiver if both are perfectly synchronized. The signal received by the receiver in the presence of a jammer is
\begin{equation}
    y[n] = y_L[n] + y_A[n],
\end{equation}
with $y_L[n]$ and $y_A[n]$ respectively the legitimate signal and the jamming signal at reception.

To simplify the notation, the received legitimate signal magnitude is assumed equal to unit, and that of the jamming signal is $V_A$, such that the jamming-to-signal amplitude ratio at the receiver is equal to $V_A$. So, a sampling window $r$ is considered jammed if the energy-dominant jamming symbol in that window has a normalized magnitude greater than 1, i.e. $\mbox{max}\big(|Y_{A}^{(r)}(k)|\big)>1$.

For the interleaving stage, we assume CR $\in \{1,2,3,4\}$ and SF $\in \{7,8,\dots,12\}$. Let $i \in \{0, 1, \dots, l-1\}$ and $j \in \{0, 1, \dots, m-1\}$. $n_s$ denotes the number of jamming symbols, where $n_s \in [0, CR+4]$.

\subsection{Synchronized jamming}
The jammer is assumed to synchronize its signal with that of the legitimate transmitter such that the first jamming symbol is assumed to be perfectly aligned in time with the first symbol of the interleaving block from the legitimate transmitter. The interleaved symbols from the legitimate signal and the jamming symbols are denoted $S^{(L)}$ and $S^{(A)}$, respectively. Specifically, $S^{(L)}{ij}$ and $S^{(A)}{ij}$ represent the $j$-th bit of the $i$-th legitimate and jamming symbols. The whitening stage allows us to assume that the transmitted symbols are random. Similarly, the jamming symbols are generated randomly, following a uniform distribution. Thus, both the jamming and legitimate symbol bits can be considered equiprobable: $p(S^{(L)}{ij}=0) = p(S^{(L)}{ij}=1) = 0.5$, and the same holds for $S^{(A)}_{ij}$. 

The probability that $S^{(L)}{ij}$ is decoded correctly (i.e., not altered by the jammer) is denoted as $p{ij}$, while $\overline{p_{ij}}$ represents the probability that it is jammed. Hence,
\vspace{-5pt}
\begin{equation}
    \begin{split}
        \overline{p_{ij}} & = P(S^{(L)}_{ij}=0) \times P(S^{(A)}_{ij}=1) \\
        & \hspace{3mm} + P(S^{(L)}_{ij}=1) \times P(S^{(A)}_{ij}=0) \\
        & = 0.5 
    \end{split}
    \label{eq:bit_jam_prob}
\end{equation}
Consequently, the jammed channel can be considered as a binary symmetric channel (BSC) with a $0.5$ error probability.
For the remainder of this paper, we define $p = p_{ij}$ and $\overline{p} = \overline{p_{ij}}$. The following analysis does not consider errors caused by noise or interference sources other than the jammer.


First, we focus on CR 1 and CR 2. Since they do not provide error correction capabilities, the frame is successfully decoded if none of the $n_s$ jamming symbols cause a bit error. Therefore, $P_{sync}$ can be expressed as:
\vspace{-5pt}
\begin{equation}
        P_{sync}(n_s)  = p^{m \times n_s}  
    \label{eq:blk_suc_prob_cr1_2}
\end{equation}

Now, we consider CR 3 and CR 4, and we define $C^{(L)}_j$ and $C^{(R)}_j$ as the $j$-th codeword of the legitimate transmitter before the interleaving stage, and the corresponding codeword at the receiver after the deinterleaving stage, respectively. Since CR 3 and 4 can correct one bit error per codeword, a block is successfully decoded if each pair $(C^{(R)}_j, C^{(L)}_j)$ has at most 1 corrupted bit. The probabilities of having 0 or 1 corrupted bit are given by Eq. \eqref{eq:ham_dist_0_1}:
\vspace{-5pt}
\begin{equation}
    \begin{split}
        Pr\left(d(C^{(R)}_j, C^{(L)}_j) = 0\right) & = p^{n_s} \mbox{    and, } \\
        Pr\left(d(C^{(R)}_j, C^{(L)}_j) = 1 \right) & = \left(n_s \atop 1 \right) p \cdot \overline{p} \; ^{n_s - 1} \\
                                                   & = n_s p \cdot \overline{p} \; ^{n_s - 1}
    \end{split}
    \label{eq:ham_dist_0_1}
\end{equation}

Therefore, the block can be correctly decoded with a probability of $P_{sync}$:
\vspace{-5pt}
\begin{equation}
    P_{sync}  = \overline{p} \; ^{m(n_s - 1)}(\overline{p} + n_s \cdot p)^m
    \label{eq:blk_suc_prb_3_4}
\end{equation}

\subsection{Non-synchronized jamming}
In reality, the legitimate signal and the jamming signal experience different propagation channels to the receiver and reach it at different times. Moreover, the response time of a jammer implemented in software is not deterministic and makes the synchronization a tedious task. Therefore, a time misalignment should be considered in the analysis. Furthermore, the jammer is assumed to be implemented with an SDR front-end with a more stable clock than that of commercial off-the-shelf LoRa transceivers. So, any frequency offsets experienced by the jamming signal can be neglected.


The probability of successfully decoding the block is:
\begin{equation}
    P_{nosync}(n_s, \tau) = p(\tau) \times P_{sync}(n_{ss}) \mbox{,}
    \label{eq:success_prob_nosync}
\end{equation}
where $p(\tau)$ is the probability of having a time offset $\tau$, $P_{sync}$ the successful decoding probability in the synchronized jamming case, and $n_{ss} = \sum_{r=0}^{n_s+1} \mathbf{1}_{\big\{\mbox{max}\big(|Y_{A}^{(r)}(k)|\big)>1\big\}}$ is the number of sampling windows successfully jammed.

Deriving a closed-form expression for $\mathbf{1}_{\big\{\mbox{max}\big(|Y_{A}^{(r)}(k)|\big)>1\big\}}$ is a tedious task as it involves the $max$ function on the spectrum of the dechirped signal. Instead, we propose to evaluate it through numerical methods over jamming signals generated with values of $\tau$ following a given distribution. Moreover, the mean of $P_{nosync}$ can be obtained by averaging over $P_{sync}(n_{ss})$ obtained for these values of $\tau$.

\subsection{Repeated-symbol jamming}
It can be shown that repeating the same symbol value for two consecutive sampling windows results in full symbol energy accumulated in one bin of the second window \cite{Hou2022_jamming_lora_phy_and_countermeasure}. When used by a jammer, this approach will successfully jam $n_s - 1$ sampling windows with full energy out of the $n_s + 1$ windows covered by the jamming signal because of the offset. The energy of the dominant bin in the first and last window will depend on $\tau$.

In this case, the probability of successful decoding is:
\begin{equation}
    \begin{split}
        P_{rep}(n_s, \tau) = & \mbox{ } p(\tau)\times P_{sync}\Big(n_s - 1 + \mathbf{1}_{\big\{\mbox{max}\big(|Y_{A}^{(0)}(k)|\big)>1\big\}} \\ 
        & + \mathbf{1}_{\big\{\mbox{max}\big(|Y_{A}^{(n_s)}(k)|\big)>1\big\}}\Big)
    \end{split}
    \label{eq:success_prob_repeat}
\end{equation}

The same evaluation method proposed for $\mathbf{1}_{\big\{\mbox{max}\big(|Y_{A}^{(r)}(k)|\big)>1\big\}}$ in Eq. \eqref{eq:success_prob_nosync} is applied here.

\subsection{Comparison of jamming approaches} 
\begin{figure}[!t]
    \centering
    \subfloat[]{\includegraphics[width=0.75\linewidth]{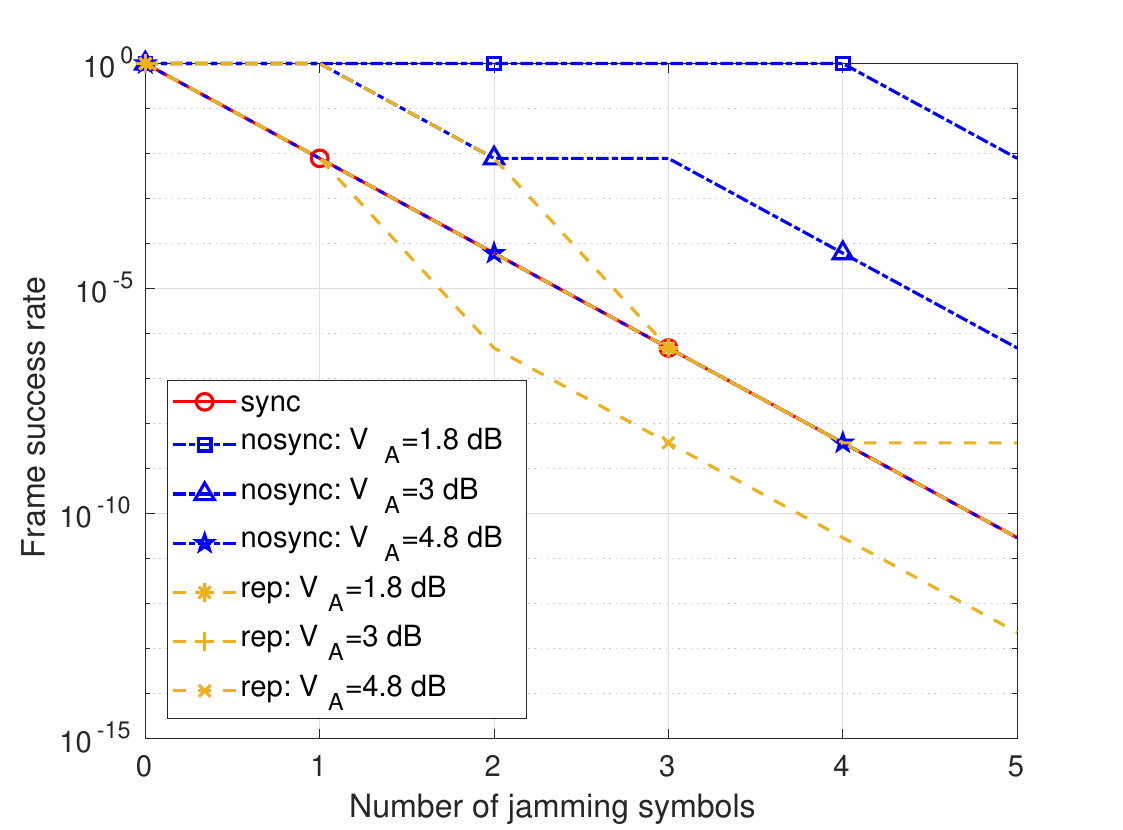}}
    \vspace{-10pt}
    
    \subfloat[]{\includegraphics[width=0.75\linewidth]{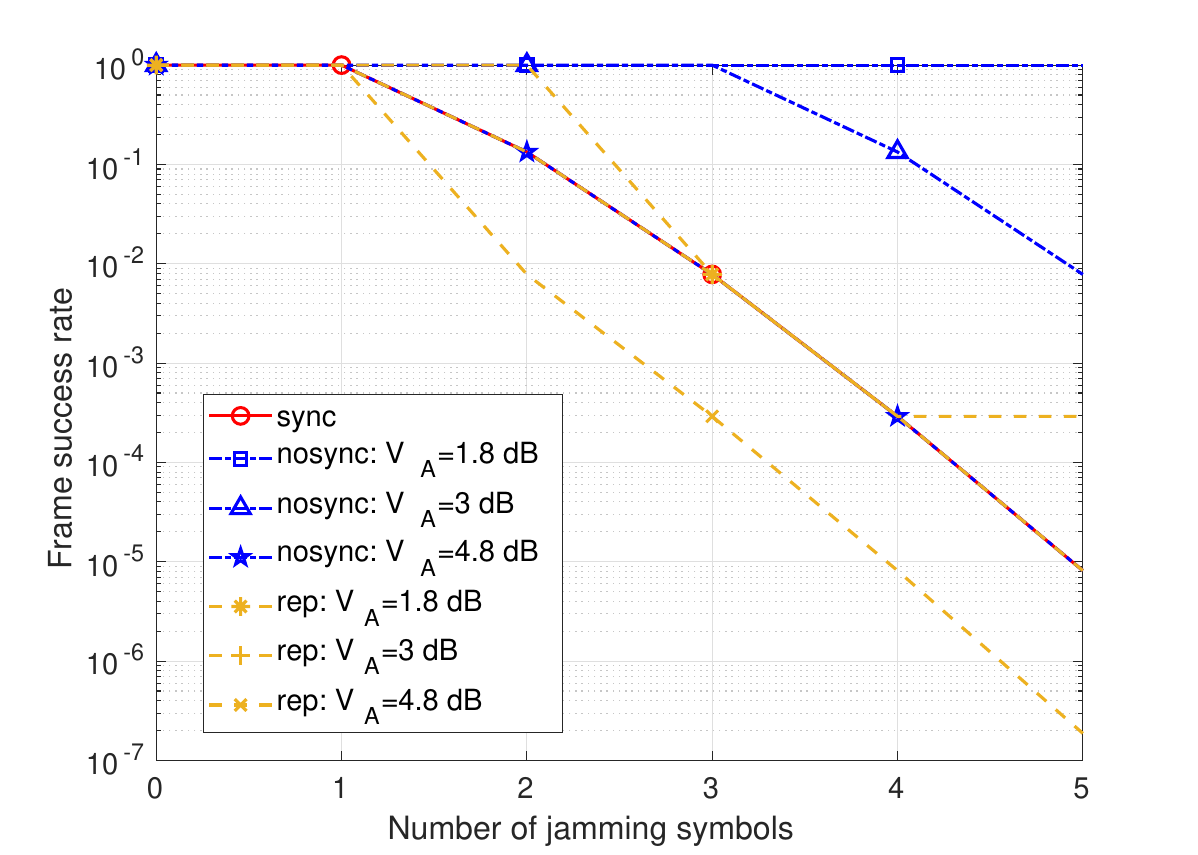}}
    \caption{Theoretical frame success rate under jamming, SF 7: (a) CR 1\&2, \mbox{(b) CR 3\&4}}
    \label{fig:frm_succ_rate_theory}
    \vspace{-18pt}    
\end{figure}

To compare the effect of the described jamming approaches on the network, the average frame success rate is plotted in Fig. \ref{fig:frm_succ_rate_theory} against jamming signal length, and different values of its magnitude $V_A$. The offset values are sampled from $\tau \sim \mathcal{N}\left( \mu = \frac{N - 1}{2},\ \sigma^2 = \left( \frac{N - 1}{6} \right)^2 \right)$, with more than $99\%$ of the values located in $[0, N-1[$. A rejection sampling is used to discard offset values outside of that interval. 

We limit our analysis to a single interleaved block for all CRs, so the maximum number of jamming symbols considered is 5. And since increasing the jamming power does not change the frame success rate for the synchronized jamming, as long as the received jamming power is superior to that of the received legitimate signal, the plot contains only one curve covering for synchronized jamming representative of all values of $V_A$.

It can be observed that the frame success is less affected by the non-synchronized jamming than it is by the synchronized jamming. And, the repeated-symbol gives the lowest frame success rate with $V_A = 4.8 \mbox{ dB}$, while it approaches the frame success of the synchronized jamming with $V_A=1.8 \mbox{ dB}$. This difference is explained by the fact that the jamming signal covers more sampling windows with the repeated-symbol jamming ($n_s+1$ compared to $n_s$ with synchronized jamming), then causes more symbol corruption at sufficiently high power. Furthermore, since most of the offsets generated following the Gaussian distribution are close to $\frac{N-1}{2}$, each non-synchronized jamming symbol is split into fractions with less power. This is, however, compensated for by the repeated-symbol jamming through the combination of the fractions of two successive symbols in a single bin. And that explains the difference in the impact of both approaches.



With CR 3\&4, the impact of the jamming is less severe because of the error-correction capability of these coding rates, even leading to a 100\% frame success rate for the non-synchronized jamming with $V_A = 1.8 \mbox{ dB}$.


\section{Experimental Assessment}
The GNU Radio implementation developed for the experiment is hosted on GitHub at \cite{lorajam_github_repo} for research purposes only, with the important disclaimer that any form of jamming is illegal. The reactive jammer is divided into three complementary blocks:
(i) The \textbf{FrameDetect} block detects a LoRa transmission, based on the central frequency, bandwidth, and SF. Upon detection, it attempts to recover its header and extract the CR. If the header is valid, it sends a message containing the SF and CR to the trigger block.
(ii) The \textbf{Trigger} block determines the number of jamming symbols to generate for each CR value. When it receives a message from the FrameDetect block, it randomly generates a symbol value, duplicates it -- repeated-jamming -- according to the prior configuration, and sends it to the jammer block. 
(iii) The \textbf{Jammer} block modulates the received symbol values according to its physical layer configuration (SF, bandwidth, sampling rate) and transmits the modulated symbols to the next block, which could be an RF transmitter.

\begin{figure}[!t]
    \centering
    \subfloat[]{\includegraphics[width=0.7\linewidth, valign=c]{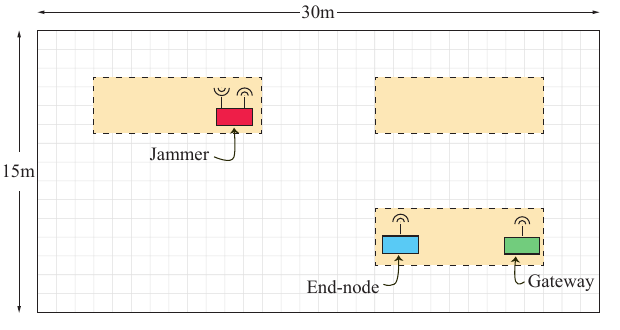} \label{fig:exp-top-view}}
    \subfloat[]{\includegraphics[width=0.25\linewidth, valign=c]{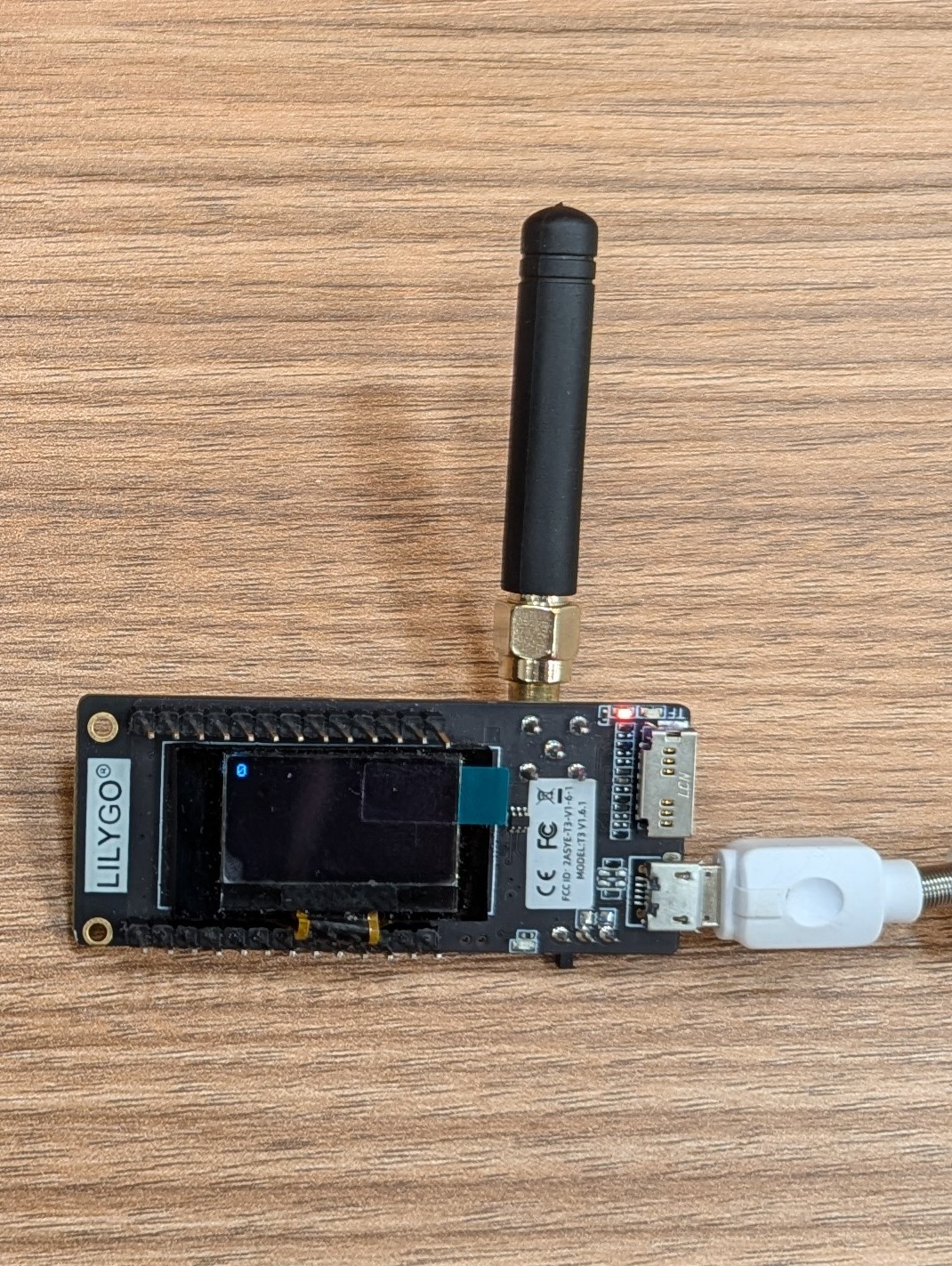} \label{fig:exp-receiver}}
    \hfill
    \subfloat[]{\includegraphics[width=0.475\linewidth]{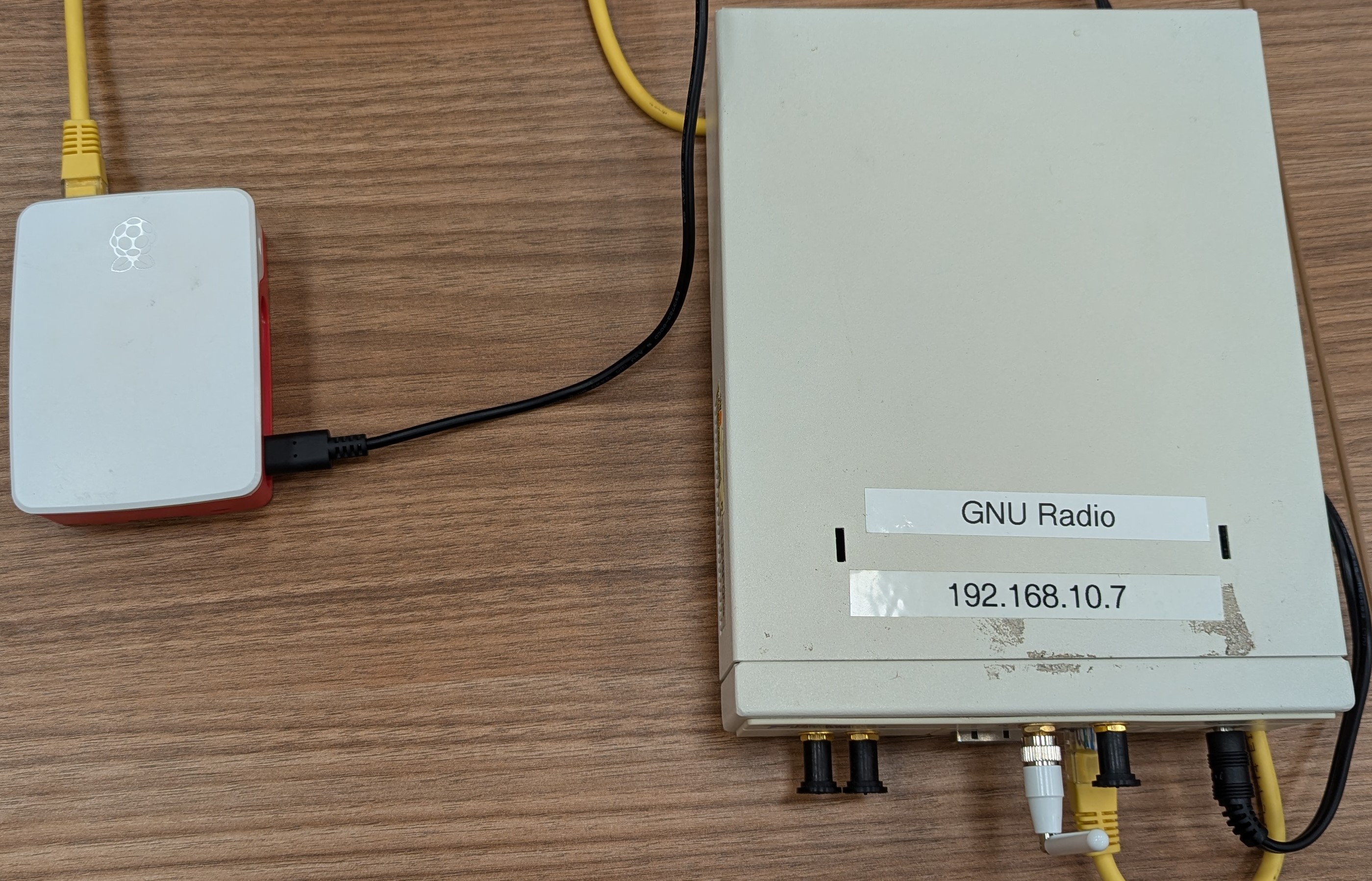} \label{fig:exp-transmitter}}
    \subfloat[]{\includegraphics[width=0.425\linewidth]{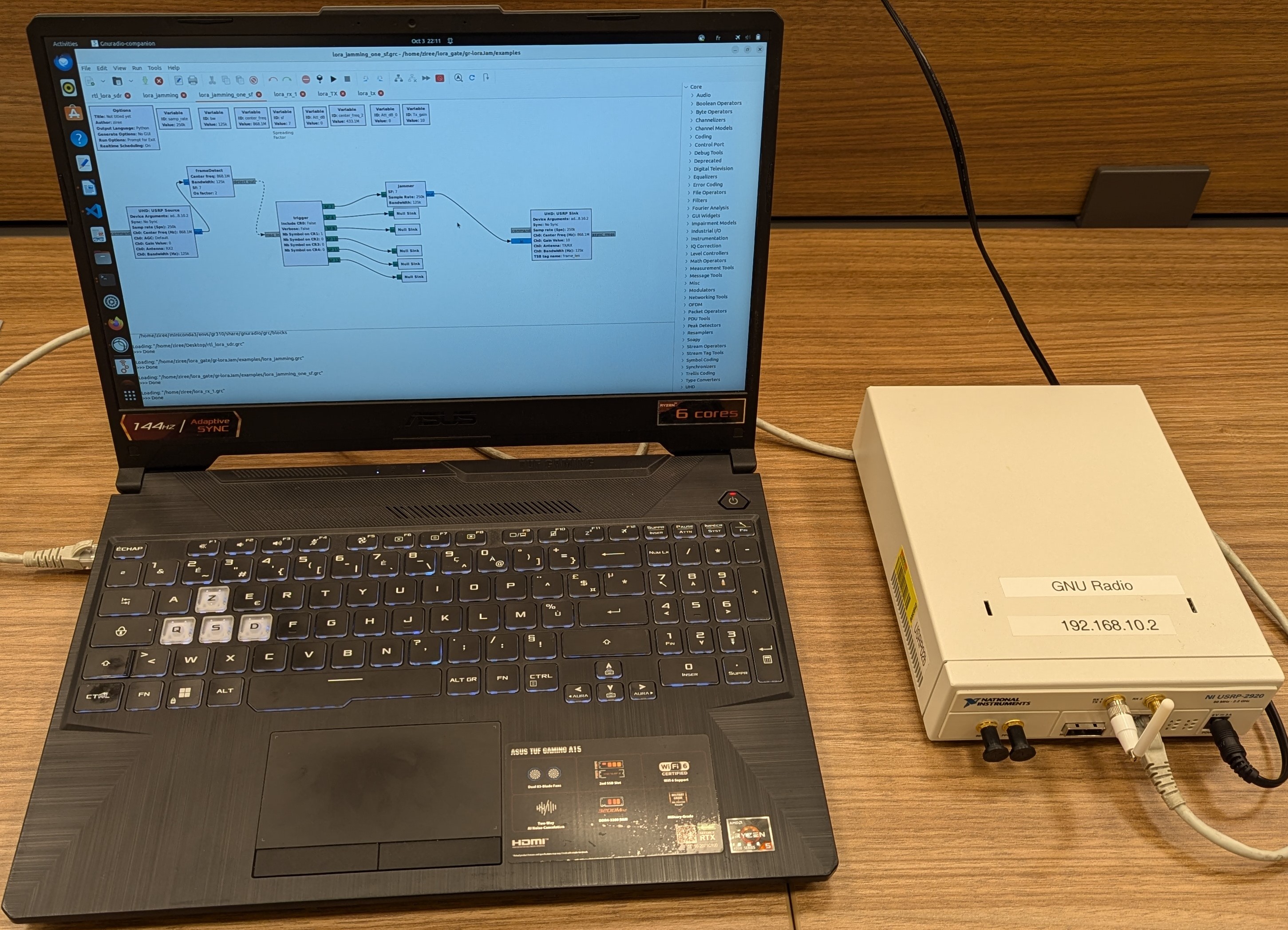} \label{fig:exp-jammer}}
    \caption{Experiment bench equipment
    : (a) Top view disposal, (b) TTGO-LoRa32 receiver, (c) USRP 2920 + Raspberry Pi 4 transmitter, (d) USRP 2920 + Laptop as Jammer
    }
    \label{fig:experiment-bench}
    \vspace{-15pt}    
\end{figure}

The experimental setup is shown in Fig. \ref{fig:experiment-bench}. The transmitting end node is a USRP-2920 connected to a Raspberry Pi 4 running the LoRa implementation from \cite{Tapparel2020_open_source_lora_phy_prototype_gnu_radio}. The jammer is a USRP-2920 connected to a laptop running our jammer implementation, and the receiver is a TTGO-LoRa32.

\begin{figure}[!t]
    \centering
    \subfloat[]{\includegraphics[width=0.745\linewidth]{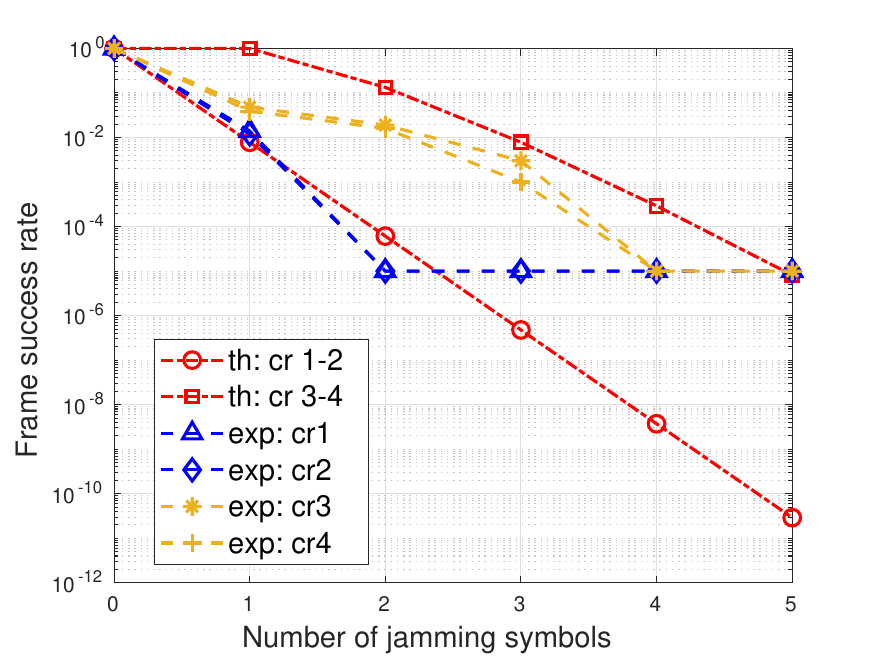} \label{fig:frm_succ_rate_th_exp_sf7}}
    \vspace{-10pt}        
    
    \subfloat[]{\includegraphics[width=0.745\linewidth]{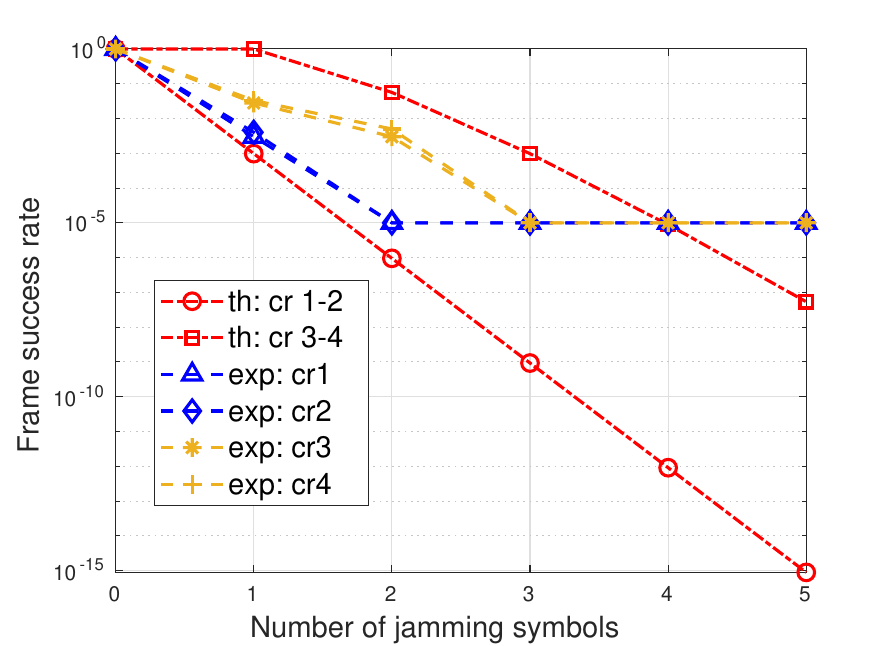}\label{fig:frm_succ_rate_th_exp_sf10}}
    \caption{Experimental Frame success rate under jamming $V_A = 3 \mbox{ dB}$: \mbox{(a) SF7, (b) SF10}}
    \label{fig:frm_succ_rate_th_exp}
    \vspace{-15pt}    
\end{figure}

The end node transmits at 10 dBm, while the jammer is configured with a transmission power of 13 dBm. The experiment considers SF 7 and SF 10, and 1000 frames are transmitted for each combination of CR and number of jamming symbols. The resulting frame success rates are displayed in Figs. \ref{fig:frm_succ_rate_th_exp_sf7} and \ref{fig:frm_succ_rate_th_exp_sf10}, along with their theoretical repeated-symbol jamming counterparts. For cases where none out of the 1000 frames were received, we set the success rate to $10^{-5}$ for better illustration.

For both SF values, the plots corresponding to CR 1 and 2 are very close to each other, as are those for CR 3 and 4. This confirms that CR 1 and 2 deliver the same performance regarding error correction, as do CR 3 and 4.

Additionally, the experimental curves closely follow the theoretical ones, with a slight deviation for CR 3\&4 and 1 jamming symbol. This difference arises from the initial assumption made on the jamming symbol offset distribution. It therefore remains an interesting task to study the real distribution of that offset.


\begin{figure}
    \centering
    \includegraphics[width=\linewidth]{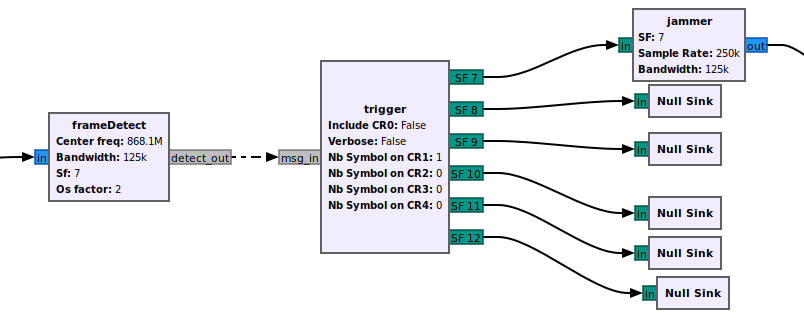}
    \caption{LoRaJam GNURadio implementation flowgraph}
    \label{fig:lorajam-gnuradio-flow}
\end{figure}

\begin{table}[!t]
    \caption{Signal to Jamming Ratio (in dB)}
    \vspace{-5pt}
    \centering
    \begin{tabular}{|c|c|c|c|c|}
    \hline
        &   CR1   &   CR2 &   CR3 &   CR4 \\
    \hline
    SF7 &   18.42 & 19.17 & 16.79 & 17.35 \\
    SF10 &  13.18 & 13.33 & 12.60 & 13.09 \\
    \hline
    \end{tabular}
    \label{tab:sjr_gain}
    \vspace{-10pt}
\end{table}

\begin{table}[!t]
    \caption{Jamming symbol gain (in dB)}
    \vspace{-5pt}
    \centering
    \begin{tabular}{|c|c|c|}
    \hline
        &   CR1\&2 &   CR3\&4 \\
    \hline
    SF7 &   9.6 & 6.59 \\
    SF10 &  9.6 & 7.84 \\
    \hline
    \end{tabular}
    \label{tab:js_gain}
    \vspace{-16pt}
\end{table}

Finally, Tables \ref{tab:sjr_gain} and \ref{tab:js_gain} demonstrate the effectiveness of our jamming approach. Table \ref{tab:sjr_gain} presents the signal-to-jamming ratio (SJR) for each combination of spreading factor (SF) and coding rate (CR). It is calculated as the ratio of the number of legitimate symbols, weighted by transmission power, to the number of jamming symbols, weighted by jamming power. The legitimate symbols correspond to physical payloads of 180 bytes for SF7 and 60 bytes for SF10, both of which are below the maximum allowable payload size. Table \ref{tab:js_gain}, on the other hand, shows the gain in jamming symbols achieved by our method compared to a jammer based on a conventional LoRa transceiver, which transmits 18.25 symbols without a payload. This once again highlights how susceptible the standard LoRa modulation is when faced with a well-designed jammer.

\section{Conclusion}
In this paper, we explored reactive jamming attacks against LoRaWAN, focusing on minimizing jammer exposure time. Our analysis demonstrates how the frame success rate is impacted by the number of jamming symbols in the case of synchronized jamming, non-synchronized jamming, and repeated-symbol jamming. The results show that transmitting fewer than one interleaving block can be enough to significantly reduce the frame success rate, provided the jamming-to-signal magnitude ratio is sufficient for the jammer’s symbol to be misinterpreted as the legitimate symbol by the receiver. Additionally, we presented a GNU Radio implementation of the jammer to validate our analysis. The deterministic and known preamble pattern of LoRa makes reactive jamming feasible and straightforward. Future work will focus on developing a stealthy preamble structure for LoRa as a countermeasure to reactive jamming.

\vspace{-0.1in}
\section*{Acknowledgment}
This work was sponsored by the Junior Faculty Development program under the UM6P-EPFL Excellence in Africa
Initiative.
\vspace{-0.05in}


\bibliographystyle{IEEEtran}
\bibliography{IEEEabrv,references}

\end{document}